\documentclass[aps,prd,preprint,superscriptaddress,tightenlines,%
nofootinbib,floatfix]{revtex4}
\newcommand{\PRE}[1]{{#1}}   

\usepackage{epsfig}
\newcommand{\postscript}[2]{\setlength{\epsfxsize}{#2\hsize}
 \centerline{\epsfbox{#1}}}

\newcommand{\ev}{\text{eV}}

\newcommand{\tev}{\text{TeV}}

\newcommand{\etal}{{\em et al.}}

\newcommand{\eqref}[1]{Eq.~(\ref{#1})}

\begin{document}

\preprint{
\hfil
\begin{minipage}[t]{3in}
\begin{flushright}
\vspace*{.4in}
NUB--3227--Th--02\\
UK/02/06\\
hep-ph/0204228
\end{flushright}
\end{minipage}
}

\title{
\PRE{\vspace*{1.5in}} Phenomenology of Randall--Sundrum Black
Holes

\PRE{\vspace*{0.3in}}
}

\author{Luis A.~Anchordoqui}
\affiliation{Department of Physics,\\
Northeastern University, Boston, MA 02115
\PRE{\vspace*{.1in}}
}

\author{Haim Goldberg}
\affiliation{Department of Physics,\\
Northeastern University, Boston, MA 02115
\PRE{\vspace*{.1in}}
}

\author{Alfred D.~Shapere}%
\affiliation{Department of Physics,\\
University of Kentucky, Lexington, KY 40502
\PRE{\vspace*{.5in}}
}


\begin{abstract}
\PRE{\vspace*{.1in}} We explore the phenomenology of microscopic
black holes in the $S^1/Z_2$ Randall-Sundrum (RS) model. We
consider the canonical framework in which both gauge and matter
fields are confined to the brane and only gravity spills into the
extra dimension. The model is characterized by two parameters,
the mass of the first massive graviton $(m_1)$, and the curvature
$1/\ell$ of the RS anti-de Sitter space. We compute the
sensitivities of present and future cosmic ray experiments to
black hole mediated events, for a wide range of $\ell$ and $m_1,$
and compare them with the sensitivities of Tevatron Runs I and II
to higher-dimensional physics. As part of our phenomenological
analysis, we examine constraints placed on $\ell$ by AdS/CFT
considerations.
\end{abstract}

\pacs{04.70.-s, 96.40.Tv, 13.15.+g, 04.50.+h}

\maketitle

One of the most exciting predictions of sub-millimeter extra
dimensions~\cite{Antoniadis:1990ew,Randall:1999ee} is the
production of black holes (BHs) in particle collisions with
center-of-mass energy larger than a TeV and sufficiently small
impact
parameter~\cite{Banks:1999gd,Emparan:2000rs,Giddings:2000ay,Giddings:2001bu,Dimopoulos:2001hw,Rizzo:2001dk}.
Although colliders have not yet attained the energies required to
probe this new strong quantum gravitational effect, the
extraordinarily high center-of-mass energies achieved at the top
of the atmosphere in ultra-high energy  cosmic ray collisions are
high enough to render any change in spacetime dimensionality
detectable~\cite{Feng:2001ib,Anchordoqui:2001ei,Emparan:2001kf,Anchordoqui:2001cg,Kowalski:2002gb}.
Among these cosmic rays, a nearly guaranteed flux of neutrinos
(produced through interactions of extremely high energy protons
with the cosmic microwave background) could produce BHs which
decay promptly initiating deeply developing air showers far above
the Standard Model (SM) rate~\cite{Feng:2001ib}, and with very
distinctive characteristics~\cite{Anchordoqui:2001ei}. In
addition, neutrinos that traverse the atmosphere  may
produce BHs through interactions in the ice or water and be
detected by neutrino telescopes~\cite{Kowalski:2002gb}. Moreover,
in scenarios with asymmetric compactifications the production of
brane configurations wrapped  around small extra dimensions may be
competitive with BH production~\cite{Ahn:2002mj}. Very recently,
based on the absence of a significant signal of deeply developing
showers reported by the AGASA Collaboration we derived new limits
on the fundamental Planck scale~\cite{Anchordoqui:2001cg}. In
this paper we expand upon this study and examine in more detail
the phenomenological implications of BH production in the
Randall--Sundrum (RS) scenario~\cite{Randall:1999ee}.

The RS model consists of two 3-branes (with equal
and opposite tensions $\sigma_{_{\rm Planck}} = - \sigma_{_{\rm SM}} =
12 \,M^3/\ell$)
which rigidly reside at $S^1/Z_2$ orbifold fixed points at the
boundaries ($y = 0$ and $y = \pi r_c$) of a slab of anti-de Sitter (AdS)
space of radius $\ell$. The classical action describing the system is given
by~\cite{Randall:1999ee}
\begin{equation}
S = M^3\, \int d^4x \int_{0}^{\pi r_c} dy \, \sqrt{-g}\;
\left( \frac{12}{\ell^2} +  R \right) \,+ \int d^4x\, \sqrt{-\eta}\,\,
\sigma_{_{\rm SM}}\,+ \int d^4x\, \sqrt{-\eta}\,\, \sigma_{_{\rm Planck}}\,,
\end{equation}
where $R$ is the 5-dimensional
Ricci scalar in terms of the metric $g_{\mu\nu}$,
$M$ is the fundamental scale of gravity, and $\eta_{ij}$ is the flat
Minkowskian metric. In what follows, the Latin
subscripts extend over ordinary 4-dimensional spacetime, whereas
Greek subscripts represent all 5 dimensions.
The metric satisfying this Ansatz (in horospherical
coordinates) reads
\begin{equation}
ds^2 = e^{-2 |y|/\ell} \,\,\eta_{ij} \,dx^i dx^j + dy^2\,.
\label{lisa-metric}
\end{equation}
Examination of the action in the 4-dimensional effective theory leads
to~\cite{Randall:1999ee}
\begin{equation}
\overline{M}_{\rm Pl}^2 = M^3\,\ell\,\, \left( 1 - e^{-2\,\pi\,r_c/\ell}\right)\,,
\label{rel}
\end{equation}
where $\overline{M}_{\rm Pl}$ is the reduced effective 4-dimensional
Planck scale. Now, assuming that SM fields are localized
on the 3-brane at $y = \pi\,r_c$, one finds that a field with the
fundamental mass parameter $m_0$ will appear to have the physical
mass $m = e^{-\pi\,r_c/\ell} \,m_0$. Hence, TeV scales can be generated from
fundamental scales of order
$M_{\rm Pl}$ through the exponential warping factor. Specifically, the
observed hierarchy between the gravitational and electroweak mass scales is
reproduced provided $r_c/\ell \approx 12$. The 4-dimensional phenomenology of this model (only gravity propagates into the bulk)
is governed by  two parameters: $c =(\ell \overline{M}_{\rm Pl})^{-1}$,
which
is expected to be near though somewhat less than unity, and $m_1$ which is
the mass of the first Kaluza--Klein graviton excitation~\cite{Davoudiasl:1999jd}.

Two different types of BHs  can be produced
in trans-Planckian particle collisions within this set up: (i)
AdS/Schwarzschild BHs that propagate freely into the bulk
(generally falling towards the AdS horizon
once produced) and (ii) tubular pancake shape BHs that are bound to the
brane~\cite{Chamblin:1999by,Giddings:2000ay}.
To study the phenomenology of
the latter, it is convenient to define a new
variable $z = \ell\,e^{y/\ell}$. In such a coordinate system
the metric
\begin{equation}
ds^2 = \frac{\ell^2}{z^2}\,\, \left(dz^2 + \eta_{ij}\, dx^i\, dx^j \right)
\end{equation}
is conformal to a 5-dimensional flat metric. To describe the SM
brane we introduce the coordinate $w = z - z_c$, where $|w| \in
(0, w_c)$, $w_c = \ell\, (e^{\pi r_c/\ell}-1)$, and the TeV brane
is located at $w=0$. After a conformal redefinition
\begin{equation}
g_{\mu\nu} \equiv \left(\frac{\ell}{z_c + w}\right)^2\tilde{g}_{\mu\nu}\,,
\end{equation}
one obtains~\cite{Hawking:73}
\begin{equation}
R = \left(\frac{z_c+ w}{\ell}\right)^2 \, \tilde{R} -
\frac{20}{\ell^2} \,,
\end{equation}
where $\tilde R$ is the Ricci scalar calculated with the metric
$\tilde g_{\mu\nu}.$ Initially Minkowskian, the metric $\tilde g$
will be modified,  under conditions to be delineated, to include
as a patch the 5-dimensional Schwarzschild solution. The gravity
sector can now be rewritten as
\begin{eqnarray}
S_{\rm gravity} &=&  M^3\,  \int d^4 x \int_0^{w_c} dw \,
\sqrt{-\tilde{g}} \; \left(\frac{\ell}{z_c+w}\right)^3 \,
\left(\tilde{R} - \frac{8}{(z_c +w)^2} \right)\nonumber\\[0.1in]
&\simeq&
  \widetilde{M}^3\, \int d^4 x \int_0^{w_{max} \ll
z_c} dw \, \sqrt{-\tilde{g}}\,\, \left(\tilde{R} -
\frac{8}{z_c^2}\right) \, ,
\label{7}
\end{eqnarray}
where
\begin{equation}
M \equiv \widetilde{M} z_c/\ell\ \ .
\label{mtilde}
\end{equation}
It is worth emphasizing that $\widetilde{M}$ refers to the
canonical frame, and is expected to be of order 1~TeV.

We will be interested in the domain of parameter space for which
a high energy collision as viewed from the SM brane can result in
the formation of a 5-dimensional spherical flat space black hole.
A calculation  of metric perturbations due to a source on the
$w=0$ brane can be made in the flat 5-dimensional space-time
approximation, {\em i.e.,} ignoring the effects of the AdS term
in Eq.~(\ref{7}), if $w\ll z_c$~\cite{Giddings:2000mu}, where $z_c$
is the AdS curvature as viewed on the SM brane, with a
gravitational constant $\widetilde M.$ This implies that to use flat
space BH formulae we must require $\tilde{r}_s \ll z_c$, where
$\tilde{r}_s$ denotes the size of 5-dimensional Schwarzschild
radius in the canonical frame~\cite{Myers:un}
\begin{equation}
\label{schwarz}
\tilde{r}_s(\widetilde{M}_{\rm BH}) = \sqrt{\frac{2\,\widetilde{M}_{\rm BH}}{3\,\pi\,\widetilde{M}_D^3}}\,,
\end{equation}
where $\widetilde{M}_{\rm BH}=\sqrt{\hat{s}}$ is the BH mass, $s$
is the center-of-mass energy in terms of $\tilde{g}_{\mu\nu}$,
and $\widetilde{M}_D = (4 \pi)^{1/3}\,
\widetilde{M}$.\footnote{We set $M_D^3 = (4 \,G_5)^{-1}$, where
$G_5$ is the 5-dimensional Newton constant.} From
Eqs.~(\ref{rel}), (\ref{mtilde}), and the forgoing relation
between $\widetilde{M}_D$ and $\widetilde M$ we find that the
condition $\tilde{r}_s < z_c$ leads to an {\em upper } bound on
the mass of the black hole for which this picture is
valid~\cite{Emparan:2001ce}:
\begin{equation}
\widetilde{M}_{\rm BH}/\widetilde{M}_D< 24\ c^{-4/3}\ \ ,
\label{upperc}
\end{equation}
where, once more, $c \equiv \left(\ell \overline{M}_{\rm
Pl}\right)^{-1} .$ When the energy exceeds this bound, the
behavior of the cross section may be analyzed within the AdS/CFT
dual picture~\cite{Maldacena:1997re}, and may assume the ln$^2E$
behavior conforming to the Froissart bound~\cite{Giddings:2002cd}.

As mentioned earlier, the parameter $c$ is expected to be
small. Comparison  of the RS brane tension and the D-brane
tension in perturbative heterotic string theory suggests that
$c\alt 0.1$~\cite{Davoudiasl:1999jd}. Work in recent years
linking the Randall-Sundrum brane-world mechanism to the
conjectured AdS/CFT correspondence~\cite{Maldacena:1997re} allows
an independent and similar bound on  $c$. The one-sided brane
world relation, Eq.(\ref{rel}) with $r_c\rightarrow \infty$, is
equivalent to
\begin{equation}
c = \left(\ell \overline{M}_{\rm Pl}\right)^{-1}=(M\ell)^{-3/2}
\ \ . \label{cml}
\end{equation}
In this ``single brane'' limit~\cite{Randall:1999vf},
Duff and Liu~\cite{Duff:2000mt} pointed out a complementarity between  AdS$_5$
and an ${\cal N}=4$ superconformal U$(N)$ Yang-Mills theory living on the
four-dimensional brane (boundary): corrections to Newton's law from both
sectors are
equal when the AdS/CFT Weyl anomaly relation~\cite{Henningson:1998gx},
\begin{equation}
2G_5N^2=\pi\ell^3\, ,
\label{weyl}
\end{equation}
or
\begin{equation}
(M\ell)^3= N^2/(8\pi^2)
\label{mln}
\end{equation}
holds. For finite but large $r_c,$ deviations from conformality are
exponentially damped in the infrared~\cite{Rattazzi:2000hs}.
BH creation in SM brane collisions becomes significant if a
weak-gravity 5-dimensional
description is valid~\cite{Arkani-Hamed:2000ds}. This occurs
for $M\ell\gg 1$, and hence $c\ll 1.$
A more quantitative estimate emerges by first combining Eqs. (\ref{cml}) and
(\ref{mln}) to obtain
\begin{equation}
c=2\sqrt{2}\pi/N\ \ .
\label{c}
\end{equation}
An approximate lower bound on $N$ may be surmised by noting that the
KK gravitons find a dual description
as glueballs resulting from strong coupling in the (approximate) CFT sector
~\cite{Arkani-Hamed:2000ds}.
In the large-$N$ analysis of 't Hooft~\cite{'tHooft:1973jz}, this
requires that the planar loop expansion parameter
$g_{\rm YM}^2N/8\pi^2>1.$ For $g_{\rm YM}^2/4\pi\simeq 0.1,$ we obtain from
(\ref{c})
\begin{equation}
c\alt 0.1 \label{cc}
\end{equation}

The key question now is at what mass ratio
$\left(\widetilde{M}_{\rm BH}/\widetilde{M}_D\right)$
is the BH description valid. According to the semiclassical
prescription, the BH evaporation is governed by its Hawking
temperature
\begin{equation}
\widetilde{T}_H = \frac{1}{2\,\pi\,\tilde{r}_s}\,.
\end{equation}
Since the wavelength $\tilde\lambda = 2\pi/\widetilde{T}_H$
corresponding to this
temperature is larger than the BH size, to a very good approximation
the BH behaves like a point-radiator with entropy
\begin{equation}
\tilde{S} = \frac{4}{3}\,\pi \,\widetilde{M}_{\rm BH} \,
\tilde{r}_s
= \sqrt{\frac{32\pi}{27}}\, \left(\frac{\widetilde{M}_{\rm
BH}}{\widetilde{M}_D}\right)^{3/2}\, \, . \label{entr}
\end{equation}
The magnitude of the entropy indicates the validity of this
picture. Thermal fluctuations due to particle emission are small
when $\tilde{S} \gg 1$~\cite{Preskill:1991tb}, and statistical
fluctuations in the microcanonical ensemble are small for $\sqrt
{\tilde S} \gg 1$~\cite{Giddings:2001bu}. In searches for BH
mediated events at colliders, it is essential to set $x_{\rm min}
\equiv \widetilde{M}_{\rm BH}^{\rm min}/\widetilde{M}_D$ high
enough that the decay branching ratios predicted by the
semiclassical picture of BH evaporation are reliable. The QCD
background is large, and therefore the extraction of signal from
background at hadron colliders depends on knowing the BH decay
branching ratios reliably. This is especially true if one is
attempting to determine discovery limits, where the overall rates
for BH production are not necessarily large.  Thus, in collider
searches, a cutoff of $x_{\rm min}  = 5.5$ (i.e., $\tilde{S} >
25$) or more may be appropriate. By contrast, the search for
deeply penetrating quasi-horizontal showers initiated by BH
decays can afford to be much less concerned with the details of
the final state, since the background is almost nonexistent.  As
a result, the signal relies only on the existence of visible
decay products, which, in this context, includes all particles
other than neutrinos, muons, and gravitons. Indeed, there is very
little about the final state, other than its total energy and to
some degree its multiplicity and electromagnetic
component~\cite{Anchordoqui:2001ei}, that we can reasonably
expect to observe, since detailed reconstruction of the primary
BH decay process is not possible at cosmic ray detectors.  It
seems reasonable to choose a significantly lower value of
$\widetilde{M}_{\rm BH}^{\rm min}$ than is needed for collider
searches; in our estimates of rates for cosmic ray facilities we
will take $x_{\rm min}$ as low as 1, or $\tilde S$ as low as 2.
While BHs of mass around $\widetilde{M}_D$ will be outside the
semiclassical regime, it seems quite reasonable to expect that
they will nevertheless decay visibly, whatever 5-dimensional
quantum gravitational description applies. Finally, as a
consequence of Eq.(\ref{upperc}) there are upper bounds on the
entropy. Combining the latter with Eq.~(\ref{entr}) we obtain
\begin{equation}
\tilde S < 250\ c^{-2}\ \ .
\end{equation}
To model the collision between two partons $i$ and $j$, with
center-of-mass energy $\sqrt{\hat s}$ and impact parameter
$\tilde{b}\alt \tilde{r_s}$, we consider a scattering amplitude
of an absorptive black disk with area
$\pi\,\tilde{r}_s^2$~\cite{Banks:1999gd}.
Criticisms~\cite{Voloshin:2001vs} of the assumptions leading to
these cross sections have been addressed~\cite{Dimopoulos:2001qe}.
Additionally,  when considering non-zero impact parameters in 5
dimensions, one must be sure that the production of BHs is favored
over that of black rings ~\cite{Emparan:2001wn} for the relevant
values of the angular momentum $J.$  For a 5-dimensional  Kerr
solution~\cite{Myers:un} with a single angular momentum $J$ (taken
in the direction of the brane~\cite{Giddings:2001bu}), the horizon
size $r_k$ and $J$ for a mass $M_{\rm BH}$ are related:
\begin{equation}
\frac{4}{9}\, r_k^2 + \frac{J^2}{M_{\rm BH}^2} =
\frac{32G_5 M_{\rm BH}}{27\pi}\ \ .
\label{rhj}
\end{equation}
A geometric cross section requires absorption in angular momenta
up to $J\simeq M_{\rm BH} r_k/2.$ On
incorporating this condition in (\ref{rhj}) one finds that the
maximum value of $J$ required for the geometric cross section is
$0.83\,M_{\rm BH}\, \sqrt{32\,G_5\,M_{\rm BH}/27\pi}$. This is
below the {\em lower} limit on the spin of a black ring, $0.92
\,M_{\rm BH}\,
\sqrt{32\,G_5\,M_{\rm BH}/27\pi}$~\cite{Emparan:2001wn}. Therefore,
it is consistent to consider
only the production of 5-dimensional BHs with geometric cross
section. For the same reasons as discussed in
\cite{Anchordoqui:2001cg}, we will calculate our cross sections
with the Schwarzschild radius.

\begin{figure}
\postscript{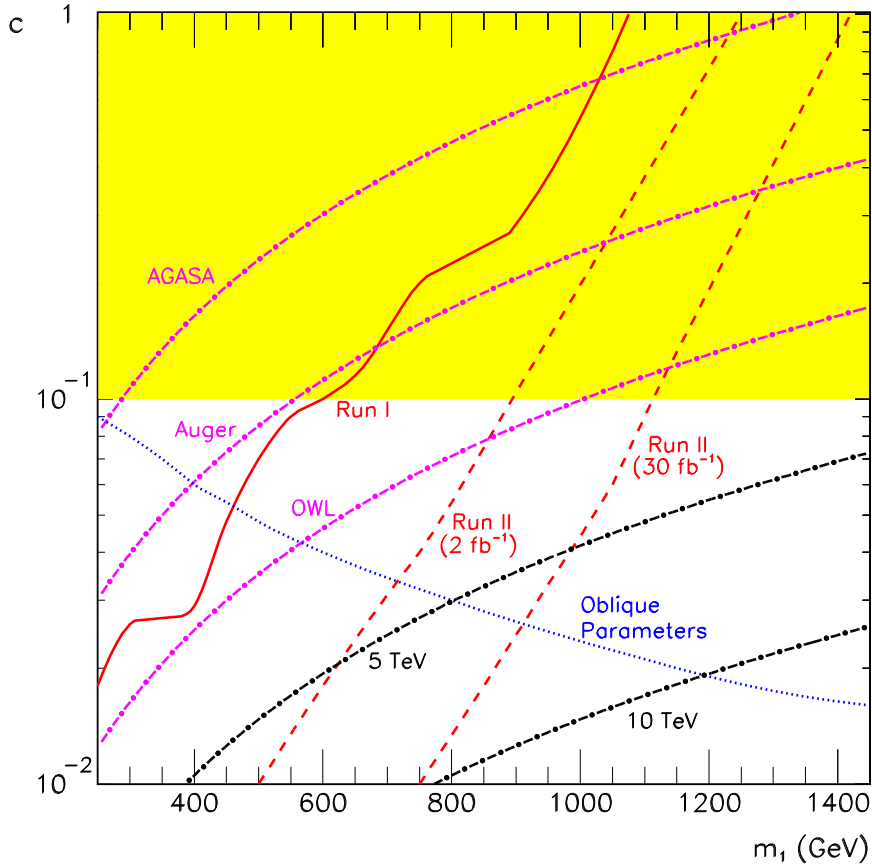}{0.80} \caption{Allowed region of
the RS parameter space. Resonant production of the first KK
graviton excitation in the Drell-Yan and dijet channels at the
Tevatron~\cite{Abachi:1996ud} has excluded the region to the left
of the solid bumpy curve~\cite{Davoudiasl:1999jd}, whereas an
analysis of the oblique parameters excludes the region below the
smoothly falling curve (dotted)~\cite{Davoudiasl:2000wi}. The
diagonal dashed lines correspond to future 95\% CL parameter
exclusion region at Run II at the Tevatron under the assumption
that no signal is found~\cite{Davoudiasl:1999jd}. The smoothly
(dashed-dotted) rising curves indicate contours of constant
$\widetilde{M}_D$ in the $c-m_1$ plane. The region above AGASA's
curve is excluded at 95\% CL. For an observation time of 5 (3)
years, Auger (OWL) will probe the region above its curve. The
unshaded area indicates the region delineated by Eq.(\ref{cc}). }
\label{nu1}
\end{figure}

In our investigation of BH production by cosmic rays, we will be most
interested in collisions of neutrinos with atmospheric nucleons.
In order to obtain the $\nu N$ cross section
we take the geometric cross section
$\hat{\sigma}$, fold in the appropriate parton densities and integrate
over the parton energies~\cite{Feng:2001ib}
\begin{equation}
\label{partonsigma}
\sigma ( \nu N \to \text{BH}) = \sum_i \int_{(\widetilde{M}_{\rm BH}^{\rm min}{})^2/s}^1 dx\,
\hat{\sigma}_i ( \sqrt{xs} ) \, f_i (x, Q) \ ,
\end{equation}
where $s = 2 m_N E_{\nu}$, the sum is over all partons in the
nucleon, and the $f_i$ are parton distribution functions.  We set
the momentum transfer $Q = \min \{ \widetilde{M}_{\rm BH},
10~\tev \}$, where the upper limit is from the CTEQ5M1
distribution functions~\cite{Lai:2000wy}. Also, as a consequence
of Eq.~(\ref{upperc}), there will be an upper bound on the parton
subenergies. However, the effect is negligible over virtually the
entire region of $c.$

With this in mind, for an incoming neutrino flux $d\Phi/dE_\nu$, the
event rate for deep showers is
\begin{equation}
{\cal N} = \int\, dE_\nu\, N_A\, \frac{d\Phi}{dE_\nu}\,
\sigma(\nu N \rightarrow {\rm BH})\,
A(E_\nu)\, T \ , \label{events}
\end{equation}
where $N_A = 6.022 \times 10^{23}$ is Avogadro's number, $A(E_\nu)$ is
the acceptance for quasi-horizontal showers in cm$^3$ water-equivalent
steradians, and $T$ is the experiment running time.
In what follows we adopt the cosmogenic neutrino flux estimates of
Protheroe and Johnson~\cite{Protheroe:1996ft}, assuming an energy cutoff
in the injection spectrum at $E_{\rm cutoff} = 3 \times 10^{21}~\ev$.
This is in very good agreement with the most recent evaluation of the
cosmogenic neutrino flux assuming a strong cosmological evolution on the
cosmic ray sources, which scales like $\propto (1+z)^4$
for redshift $z<1.9$ and becomes flat at higher
redshifts~\cite{Engel:2001hd}. To explore possible
additional contributions from semi-local nucleon sources, we also
consider below the cosmogenic neutrino flux estimates of Hill and
Schramm~\cite{Hill:1985mk}. Bounds on $\widetilde{M}_D$ emerge as a result of
adopting the AGASA limit of 1 possible event observed, with a
background of 1.72 events~\cite{agasa}. This places a limit of $\le
3.5$ events at 95\% CL~\cite{Feldman:1997qc}. Inserting our cross
section into \eqref{events} and requiring ${\cal N}\le 3.5$ leads
to exclusion of the region
$ \widetilde{M}_D \alt 0.70\pm 0.05~{\rm TeV}$ at 95\% CL~\cite{Anchordoqui:2001cg}.
The variation in this bound reflects uncertainty in the neutrino flux.
The forthcoming facilities of the Auger Observatory~\cite{Auger} will
probe fundamental
Planck scales, $\widetilde{M}_D\alt 1.55\pm
0.15~{\rm TeV}$~\cite{Anchordoqui:2001cg}, assuming 5
years of data and no excess above SM neutrino background. The variation in
this case includes  uncertainties in the hadronic background and the
experimental aperture.

Beyond Auger, NASA has authorized studies for the satellite
cosmic ray detection facility known as the Orbiting Wide-angle
Light-collectors (OWL)~\cite{OWL}, projected for 2007. In a
comprehensive recent study, BH event rates at OWL for $n\ge 2$
large extra dimensions were reported~\cite{Dutta:2002ca}. The
greatly increased aperture of this space-based facility will
allow the detection $\sim 10$ BH/yr for $x_{\rm min}$ as large as
5. This is well into the region of large entropy, permitting
comparison with characteristic shower profiles. In the case of
present interest $(n=1),$ the ``eyes of the OWL'' will
substantially extend the region of the $c-m_1$ plane probed in
cosmic ray observations. Under the assumption that the cosmic ray
sources have a strong cosmological evolution, the expected
background from SM neutrino interactions (all flavors) is 3
events/year~\cite{Dutta:2002ca}. On the basis of zero hadronic
background, three years of observation  implies a deviation from
the SM at a 95\% CL for 7.77 events observed above
background~\cite{Feldman:1997qc}. If these are BHs produced with
$x_{\rm min}=1$, this represents a sensitivity to
$\widetilde{M}_D\alt 2.8$~TeV at the 95\% CL~\cite{Ina}.

\medskip
\noindent{\em Summary.}  The  Randall-Sundrum model predicts a
series of TeV-scale graviton resonances  with weak scale
couplings to SM fields: this makes the Tevatron an outstanding
probe of RS physics. Analysis~\cite{Davoudiasl:1999jd} of Tevatron
data~\cite{Abachi:1996ud} for anomalous Drell-Yan and dijet
production as well as the calculation of indirect contributions
to electroweak observables~\cite{Davoudiasl:2000wi} already place
significant constraints  on the $c-m_1$ plane. In this work we
have developed the formalism for calculating RS black hole
production through cosmic ray neutrino collisions in the
atmosphere, and have delineated additional constraints placed on
the RS model by existing and future cosmic ray data. The complete
allowed region is summarized in Fig.~1. The lines of constant
$\widetilde{M}_D$ reflect the relation $m_1=3.83\ c^{2/3}\
(4\pi)^{-1/3} \widetilde{M}_D.$ AGASA is able only marginally to
probe the parameter space allowed by the Tevatron in the limited
region $m_1\agt 1$~TeV and $c > 0.6 - 0.7$. Five years of running
at Auger can reveal events in the region $m_1 > 1$~TeV and
$c>0.5$ which will not be tested by the Tevatron Run II A. A year
of observation by OWL will probe a substantially larger part of
the parameter space for small $c$ (see Fig. 1). We have also
delineated in our figure the range of the parameter $c$ (related
to the AdS radius) favored through AdS/CFT considerations.

\begin{acknowledgments}
We would like to thank Jonathan Feng for continuing encouragement
and enlightening discussions. The work of LAA and HG has been
partially supported by the US National Science Foundation (NSF),
under grants No.\ PHY--9972170 and No.\ PHY--0073034,
respectively. The work of ADS is supported in part by the
Department of Energy (DOE) Grant No.\ DE--FG01--00ER45832 and NSF
Grant No.\ PHY--0071312.
\end{acknowledgments}

\end{document}